\documentclass[aps,prc,twocolumn,superscriptaddress,showpacs,floatfix,nofootinbib]{revtex4-1}
\usepackage{amsmath,graphicx}

\begin{document}

\title{Breaking of factorization of two-particle correlations in hydrodynamics}

\author{Fernando G.  Gardim}
\author{Fr\'ed\'erique Grassi}
\affiliation{
Instituto de F\'\i sica, Universidade de S\~ao Paulo, C.P. 66318, 05315-970, S\~ao Paulo-SP, Brazil}
\author{Matthew Luzum}
\affiliation{
CNRS, URA2306, IPhT, Institut de physique th\'eorique de Saclay, F-91191
Gif-sur-Yvette, France}
\affiliation{
McGill University,  3600 University Street, Montreal QC H3A 2TS, Canada
}
\affiliation{
Lawrence Berkeley National Laboratory, Berkeley, CA 94720, USA
}
\author{Jean-Yves Ollitrault}
\affiliation{
CNRS, URA2306, IPhT, Institut de physique th\'eorique de Saclay, F-91191
Gif-sur-Yvette, France}

\date{\today}

\begin{abstract}
The system formed in ultrarelativistic heavy-ion collisions behaves as
a nearly-perfect fluid. This collective behavior is probed 
experimentally by two-particle azimuthal correlations, which are 
typically averaged over the properties of one particle in each 
pair. 
In this Letter, we argue that much additional information is contained
in the detailed structure of the correlation. 
In particular, the correlation matrix exhibits an approximate 
factorization in transverse momentum, which is taken as a strong evidence for the
hydrodynamic picture, while deviations from the factorized form  are 
taken as a signal of intrinsic, ``nonflow'' correlations. 
We show that hydrodynamics in fact predicts factorization breaking
as a natural consequence of initial state fluctuations and averaging
over events. 
We derive the general inequality relations that hold if flow dominates,
and which are saturated if the matrix factorizes. 
For transverse momenta up to 5~GeV, these inequalities are satisfied 
in data, but not saturated.  
We find factorization breaking in 
event-by-event ideal hydrodynamic
calculations that is at least as large as in data, and argue that this phenomenon opens a new 
window on the study of initial fluctuations. 
\end{abstract}


\maketitle

\section{Introduction}
In relativistic heavy-ion collision experiments a large second Fourier harmonic is observed in two-particle correlations as a function of relative azimuthal angle~\cite{Ackermann:2000tr,Adler:2003kt,Back:2004mh,Aamodt:2010pa}.  This has long been considered a sign of significant collective behavior~\cite{Ollitrault:1992bk}, or ``elliptic flow'', indicating the existence of a strongly-interacting, low-viscosity fluid~\cite{Romatschke:2007mq}.  However, only recently has it been realized that \textit{all} such correlations observed between particles separated by a large relative pseudorapidity could be explained by this collective behavior~\cite{Mishra:2007tw, Sorensen:2008dm,Takahashi:2009na,Hama:2009vu,Andrade:2010xy,Sorensen:2010zq,Alver:2010gr,Staig:2010pn,Luzum:2010sp}, at least for the bulk of the system.

One significant piece of evidence for this view was the recent observation of the factorization~\cite{Aamodt:2011by,Alver:2010rt,Chatrchyan:2012wg,ATLAS:2012at} of two-particle correlations into a product of a function of properties of only one of the particles times a function of the properties of the second.  Specifically, for pairs of particles in various bins of transverse momentum $p_T$, factorization of each Fourier harmonic was tested as~\cite{Aamodt:2011by}:
\begin{equation}
\label{test}
V_{n\Delta} (p_T^a, p_T^b) \equiv \left\langle \cos n(\phi^a - \phi^b) \right\rangle \overset{?}{=} v_n (p_T^a) \times v_n (p_T^b) ,
\end{equation}
where the brackets indicate an average over pairs of particles ($a$ and $b$) coming from the same event as well as an average over a set of collision events, and $\phi^a(\phi^b)$ is the azimuthal angle of particle $a(b)$.  The left-hand side is a (symmetric) function of two variables, $p_T^a$ and $p_T^b$, and in general may not factorize into a product of a function $v_n$ of each variable individually.   The fact that this factorization holds at least approximately, then, is a non-trivial observation about the structure of the correlation.

While most known sources of non-flow correlations do not factorize at low $p_T$~\cite{Kikola:2011tu}, a type of factorization comes naturally in a pure hydrodynamic picture where particles are emitted independently.  They thus have no intrinsic correlations with other particles, carrying only information about their orientation with respect to the system as a whole.  This causes the two-particle probability distribution in a single collision event to factorize~\cite{Dinh:1999mn} into a product of one-particle distributions,
\begin{equation}
\label{hydro}
\frac {dN_{pairs}}{d^3p^a d^3p^b}  \overset{\rm{(flow)}}{=} \frac {dN}{d^3p^a}\times \frac {dN}{d^3p^b}.
\end{equation}

Inspired by this fact, it has often been stated~\cite{ATLAS:2012at,Adare:2011hd,collaboration:2011hfa} that the factorization test in Eq.~\eqref{test} should work perfectly in hydrodynamics.  The observed approximate factorization was hailed as a success for the flow interpretation of correlations, while small deviations from the factorized form was interpreted as a gradual breakdown of the hydrodynamic description with increasing transverse momentum, and of increasing contribution from other sources of correlations.

In this work, we show that factorization as in Eq.~\eqref{test} is not
necessarily present even in an ideal hydrodynamic system governed by
Eq.~\eqref{hydro}, because of event-by-event 
fluctuations~\cite{Miller:2003kd,Alver:2006wh,Alver:2010gr}. 
These stem from quantum fluctuations:
the collision takes place over a very short timescale, and takes a
snapshot of the wavefunction of incoming nuclei. 
In the presence of fluctuations, 
we show that the correlation matrix satisfies general inequalities,
which are saturated by Eq.~\eqref{test}. 
We test these inequalities on ALICE data and point out where 
breaking of factorization occurs. 
We then illustrate with a full event-by-event hydrodynamic calculation
that the same deviation seen in experiment is also present in ideal
hydrodynamics. 


\section{Hydrodynamics and two-particle correlations}
We begin by recalling the discussion originally found in Ref.~\cite{Luzum:2011mm}.   In a pure hydrodynamic picture, particles are emitted independently from the fluid at the end of the system evolution according to some underlying one-particle probability distribution.  One can write any such distribution as a Fourier series in the azimuthal angle $\phi$ of the particles
\begin{equation}
\label{complexfourier}
 \frac {2\pi} {N} \frac {dN}{d\phi} = \sum_{n=-\infty}^{\infty} V_n(p_T,\eta) e^{-i n \phi},
\end{equation}
where $V_n = \{e^{in\phi}\}$ is the $n$th complex Fourier flow
coefficient, and curly brackets indicate an average over the
probability density in a single event.  
Writing $V_n=v_ne^{in\Psi_n}$, where $v_n$ is the (real) anisotropic
flow coefficient and $\Psi_n$ the corresponding phase, and using $V_{-n}=V_n^*$
(where $V_n^*$ is the complex conjugate of $V_n$), this can be
rewritten as 
\begin{equation}
\label{realfourier}
 \frac {2\pi} {N} \frac {dN}{d\phi} = 1 + 2\sum_{n=1}^\infty
 v_n(p_T,\eta) \cos n\left(\phi - \Psi_n(p_T,\eta)\right).
\end{equation}
Note that, for this form to describe an arbitrary distribution, both $v_n$ and $\Psi_n$ may depend on transverse momentum $p_T$ and pseudorapidity $\eta$.

In this picture, the relation in Eq.~\eqref{hydro} holds, and a complex Fourier harmonic of the two-particle correlation factorizes in each event as:
\begin{eqnarray}
 \label{complex} 
\left\{ e^{in(\phi^a - \phi^b)} \right\} &=&
\left\{e^{in\phi^a}\right\} \left\{ e^{-in\phi^b} \right\}\cr
 &=& V_n^{a}V_n^{b*}=v_n^a v_n^b e^{in (\Psi_n^a - \Psi_n^b)}.
\end{eqnarray}
This factorization only holds in a single hydro event. Both the
magnitudes and the phases of anisotropic flow fluctuate
event to event~\cite{Miller:2003kd,Alver:2006wh,Alver:2010gr}. 
The experimental quantity, Eq.~\eqref{test}, is then obtained by
averaging over events:
\begin{equation}
\label{complexav}
V_{n\Delta} (p_T^a, p_T^b) =\left\langle V_n^{a}V_n^{b*}\right\rangle=
\left\langle v_n^a v_n^b e^{in (\Psi_n^a -
    \Psi_n^b)}\right\rangle 
\end{equation} 
Due to parity symmetry, only the real part remains after this average, hence the cosine in Eq.~\eqref{test}.

From this relation alone, one can make the following general
statements about the event-averaged correlation matrix: the diagonal elements must be positive, and the off-diagonal elements must satisfy a Cauchy-Schwarz inequality,
\begin{align}
V_{n\Delta}(p_T^a,p_T^a) &\geq 0 ,\\
\label{CS}
V_{n\Delta}(p_T^a,p_T^b)^2 &\leq V_{n\Delta}(p_T^a,p_T^a)V_{n\Delta}(p_T^b,p_T^b) .
\end{align}
Factorization, Eq.~\eqref{test}, implies that the second inequality is
saturated, i.e., equality is achieved. 
Thus, while flow does not necessarily imply factorization, 
any violation of these inequalities is an unambiguous indication of 
the presence of non-flow correlations.

An inspection of published data from the ALICE
Collaboration~\cite{Aamodt:2011by} shows that these inequalities are
indeed violated in certain regimes~\cite{Ollitrault:2012cm}. 
For $n=3$, diagonal elements $V_{3\Delta}(p_T^a,p_T^a)$
are negative above 5~GeV for 0--10\% centrality, and above 4~GeV for
40--50\% centrality. 
This is a clear indication that there are nonflow correlations at high
$p_T$. For instance, the correlation between back-to-back jets
typically yields a relative angle $\Delta\phi\sim\pi$, thus 
producing a negative $V_{3\Delta}$ at high $p_T$. 
For $n=1$, diagonal elements are negative not only at high $p_T$ (with a
slightly higher threshold than for $n=3$), but also for $p_T$ between 1
and 1.5~GeV. This is believed to be caused by the correlation from
global momentum conservation~\cite{Retinskaya:2012ky,ATLAS:2012at}, but it is
interesting to note that its effect can be noticed by a simple
inspection of elements. 

\begin{figure*}
\includegraphics[width=\linewidth]{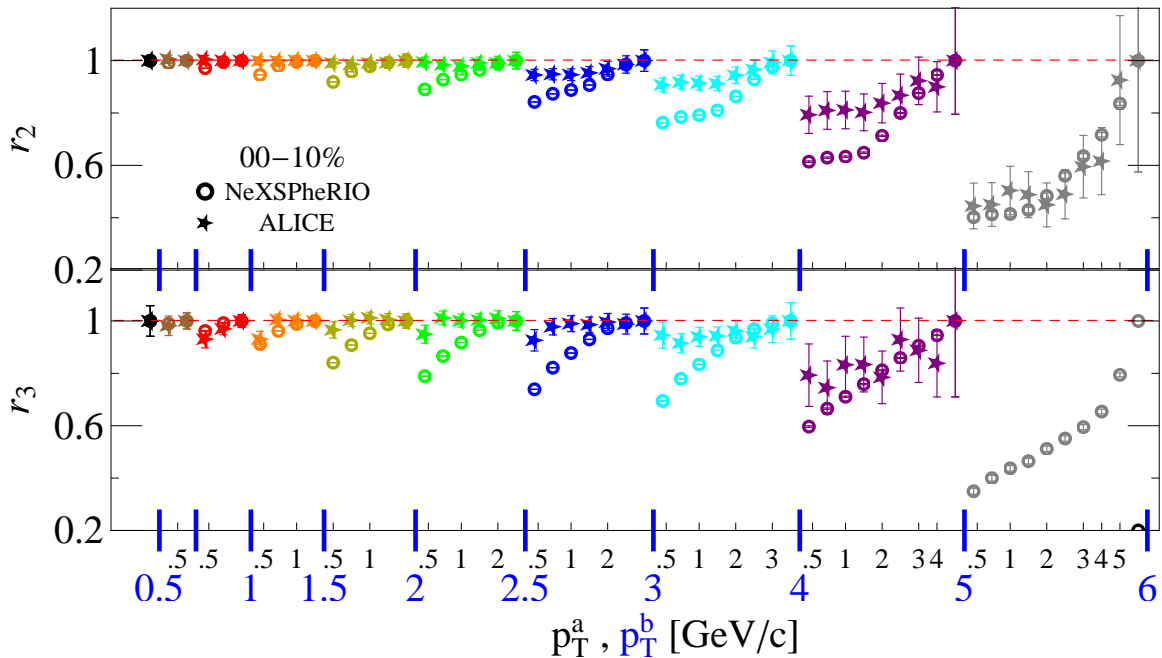}
\caption{(Color online) 
The ratio of nondiagonal to diagonal correlations, defined by
Eq.~\eqref{correlation}, is plotted on an interleaved $p_T^a$, $p_T^b$ axis; $p_T^b$ is constant between each long hash on the x-axis.
Filled stars: ALICE data for 0-10\%  central Pb-Pb
collisions at 2.76~TeV~\cite{Aamodt:2011by}. 
Open cicles: ideal hydrodynamic calculations for 0-10\%
central Au-Au collisions at 200~GeV. 
}
\label{fig:cauchy}
\end{figure*}

In order to check the validity of the second inequality \eqref{CS}, we
introduce the ratio

\begin{equation}
\label{correlation}
r_n\equiv\frac{V_{n\Delta}(p_T^a,p_T^b)}{\sqrt{V_{n\Delta}(p_T^a,p_T^a)V_{n\Delta}(p_T^b,p_T^b)}}, 
\end{equation}
 
which is defined when diagonal elements $V_{n\Delta}(p_T^a,p_T^a)$ and $V_{n\Delta}(p_T^b,p_T^b)$ are both positive, and lies between $-1$
and $+1$ if Eq.~\eqref{CS} holds. 
Factorization corresponds to the limit $r_n=\pm 1$. 
Figure~\ref{fig:cauchy} displays $r_2$ and $r_3$ as a function of $p_T^a$ and
$p_T^b$ for Pb-Pb collisions at 2.76~TeV, 0--10\%
centrality. 
ALICE results for $r_2$ satisfy the inequalities \eqref{CS} at all
$p_T$. When both particles are below $1.5$~GeV, the inequality is saturated, $r_2=1$, within errors. 
As soon as one of the particles is above $1.5$~GeV, however, $r_2$ is
smaller than unity, and the difference with unity increases with the
difference $p_T^a-p_T^b$. 
Results for $r_3$ are qualitatively similar below 5~GeV, with 
larger error bars. However, $r_3$ is closer to 1 than $r_2$ between
2 and 3~GeV. 
The values of $r_n$ for mid-central collisions (40-50\% centrality, not
shown) are comparable to the values for central collisions, although 
$r_2$ is slightly closer to 1. 

The ALICE collaboration concluded from their analysis that
factorization holds approximately for $n>1$ and $p_T$ below $4$~GeV. 
However, their results actually show evidence for a slight breaking of
factorization for $n=2$, as soon as one of the particles has $p_T>1.5$~GeV. 
Even though factorization is broken, the general inequalities implied
by flow are satisfied for $n=2$ and $n=3$ below 5~GeV for central collisions. 
It is therefore worth investigating in more detail to what extent  the
breaking of factorization which is seen experimentally can be
understood within hydrodynamics. 

First, we recall under which conditions factorization holds in
hydrodynamics. It implies that the Cauchy-Schwarz
inequality~\eqref{CS} is saturated. 
By inspection of Eq.~\eqref{complexav}, this 
in turn implies that the complex flow vectors $V_n^a$ and
$V_n^b$ are linearly dependent. 
This is true only under the following assumptions:

\begin{enumerate}
\item
By parity symmetry, $\Psi_n^a-\Psi_n^b = 0$ in each event. I.e.,
$\Psi_n$ does not depend on $p_T$, which removes the exponential from
the right-hand side of Eq.~\eqref{complex}. 
\item
$v_n(p_T)$ changes from event to event by only a
global factor, with no $p_T$-dependent fluctuations. 
$v_n(p_T)$ in the right-hand side of Eq.~\eqref{test} then represents
the rms value over events.
\end{enumerate}

In general, fluctuations ensure that these conditions are not met
exactly, and the factorization of Eq.~\eqref{test} will not be
perfect. 
Within hydrodynamics, the ratio $r_n$ in Eq.~\eqref{correlation} has a
simple interpretation. Inserting Eq.~\eqref{complexav} into
Eq.~\eqref{correlation}, one obtains

\begin{equation}
\label{rhydro}
r_n=\frac{\langle V_n^{a*}V_n^{b}\rangle}{\sqrt{\langle |V_n^a|^2\rangle
\langle |V_n^b|^2\rangle}},
\end{equation}

The ratio $r_n$ thus represents the linear correlation between the complex flow vectors at
momenta $p_T^a$ and $p_T^b$. Since in each event, $V_n^a$ is a smooth function of $p_T^a$, one
expects that the correlation is stronger when $p_T^a\simeq p_T^b$, and
decreases as the difference between $p_T^a$ and $p_T^b$ increases, as
a result of the decoherence induced by initial fluctuations. 
ALICE data confirm this qualitative expectation. 

Note that even in a single hydrodynamic event, factorization holds in
the complex form Eq.~\eqref{complex}, but is broken if one takes the 
real part before averaging over particle pairs, as in
Eq.~\eqref{test}. The ratio $r_n$ in Eq.~\eqref{rhydro} is then $\cos n(\Psi_n^a-\Psi_n^b)$, which is smaller than unity as soon as the flow
angle $\Psi_n$ depends on $p_T$. 

The question then becomes:  how large are factorization-breaking
effects in hydrodynamics, and do they have the same properties as seen in data?  If purely hydrodynamic calculations give the same result as experiment, then the observed breaking of factorization may not indicate the presence of non-flow correlations.


\section{Ideal hydrodynamic calculations}

To illustrate these concepts we perform calculations using the
NeXSpheRIO model~\cite{Hama:2004rr}. 
This model solves the equations of relativistic ideal hydrodynamics
with fluctuating initial conditions given by the NeXuS event
generator~\cite{Drescher:2000ha}. 
It has proven succesful in reproducing RHIC results, in particular the
structure of two-particle angular correlations in Au-Au 
collisions at the top RHIC energy~\cite{Takahashi:2009na}. 
It has recently been shown to reproduce the whole set of measured
anisotropic flow data~\cite{Gardim:2012yp,Gardim:2011qn,DerradideSouza:2011rp}. 
Our calculations are therefore performed for Au-Au collisions at the
top RHIC energy, not for Pb-Pb collisions at LHC energy, as would be
appropriate for a direct quantitative comparison with ALICE data. 
Our results are merely meant as a proof of concept, and as a 
prediction for measurements at RHIC. 
Note that the main source of fluctuations (namely, the finite number of
nucleons within the nucleus) is identical in both cases.

We run $30000$ NeXuS events, which are then sorted into $10\%$
centrality bins defined by the number of participant nucleons, and 
then evolved hydrodynamically. 
Anisotropic flow is calculated accurately in every event~\cite{Gardim:2011xv}. 
The ratio $r_n$ is displayed in Fig.~\ref{fig:cauchy} for $n=2$ and
$n=3$. Deviations from the factorization limit $r=1$ are already seen
at low momentum but become larger as the difference between $p_T^a$
and $p_T^b$ increases, as expected from the general arguments above. 
Surprisingly, the breaking of factorization appears {\it
  larger\/} in hydrodynamics than in experiment. 

\begin{figure*}
\includegraphics[width=\linewidth]{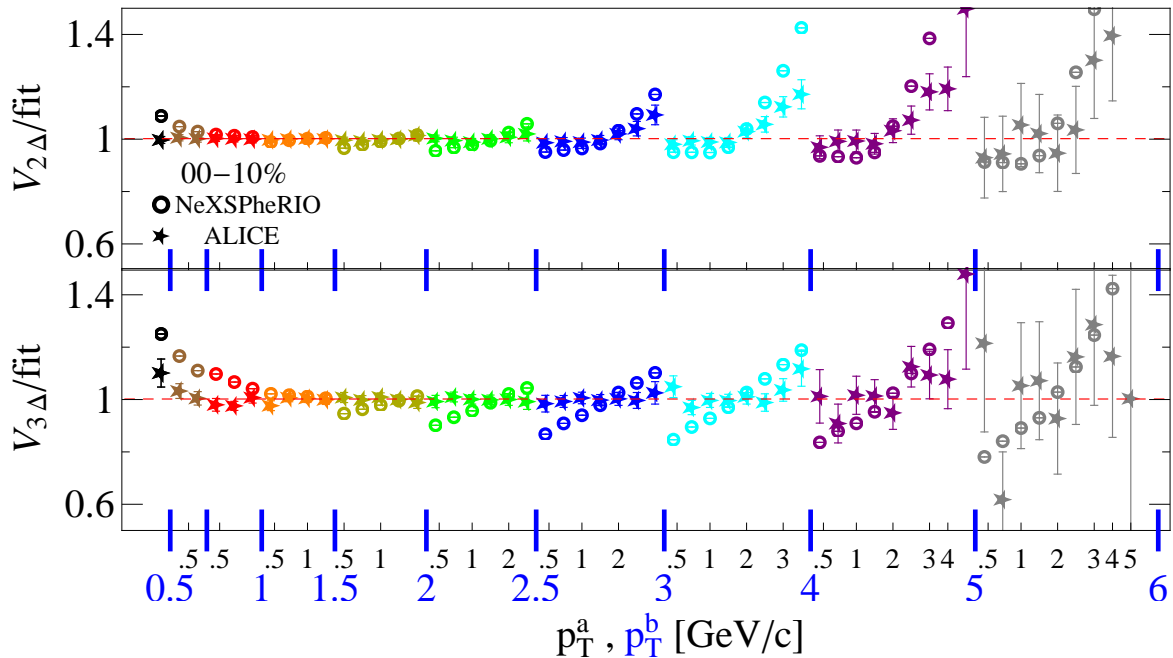}
\caption{(Color online) 
Ratio of the left-hand side to the right-hand side of
Eq.~\eqref{test}. 
Filled stars: ALICE data for 0-10\%  central Pb-Pb
collisions at 2.76~TeV~\cite{Aamodt:2011by}. 
Open cicles: ideal hydrodynamic calculations for 0-10\%
central Au-Au collisions at 200~GeV. 
}
\label{fig:globalfit}
\end{figure*}
The ALICE collaboration has studied factorization by performing a
global fit of the measured correlation $V_{n\Delta}(p_T^a,p_T^b)$ by the
right-hand side of Eq.~\eqref{test}, where $v_n(p_T)$ is a fit
parameter~\cite{Aamodt:2011by}.  
The ratio of the measured correlation 
to the best fit differs from unity if factorization is broken.
We can apply the same procedure to our hydrodynamic results. 
The result is shown in Fig.~\ref{fig:globalfit}. 
Again, hydrodynamic calculations and experimental data show similar
trends, with the noticeable difference that the breaking of
factorization is significantly {\it stronger\/} in ideal hydrodynamics
than in data. 

Several effects can explain this discrepancy. First, the average $p_T$
is significantly larger at LHC than at RHIC~\cite{Floris:2011ru},
so that it might be more natural to compare, e.g.,
4~GeV at RHIC to 5~GeV at LHC, rather than doing the comparison at the
same $p_T$. 
The second effect is viscosity, which is neglected in our
calculation. Shear viscosity, in particular, tends to damp the effect
of initial fluctuations~\cite{Schenke:2010rr}. It
is therefore natural that it will also decrease the 
breaking of factorization induced by initial fluctuations.  
A similar observation is that the linear correlation between the
initial eccentricity and the final anisotropic flow is stronger in
viscous hydrodynamics~\cite{Niemi:2012aj} than in ideal
hydrodynamics~\cite{Gardim:2011xv}.  

\section{Conclusions}

We have demonstrated that the detailed structure of two-particle
angular correlations contains much more information than traditional
analyses of anisotropic flow, where the correlation is averaged over
one of the particles~\cite{Luzum:2012da}.
Even though such two-dimensional analyses are much more demanding
in terms of statistics than traditional analyses, they bring new, independent
insight into the underlying physics of flow fluctuations. 

In particular, we have shown that quantum fluctuations in the
wavefunction of incoming nuclei result in a decoherence in the angular
correlations produced by collective flow, which becomes increasingly
important as the difference between particle momenta increases. 
Due to this effect, factorization of angular correlations is 
broken even if collective flow is the only source of correlations. 
Our numerical calculations show that factorization breaking can be as 
strong in hydrodynamics as in experimental data,
thereby suggesting that all correlations below $p_T\sim 5$~GeV (for
central Pb-Pb collisions at LHC near midrapidity) may actually be
dominated by flow.  
The sensitivity of this decoherence phenomenon to viscosity has
not yet been investigated, but we anticipate that factorization should
be restored as viscosity increases, thus potentially offering a new
means of constraining the viscosity from data. 
On the other hand, thermal fluctuations should be considered along with
viscosity~\cite{Kapusta:2011gt}, and may also contribute to
factorization breaking. 

Decoherence also provides a natural explanation for the important observation that 
event-by-event fluctuations reduce elliptic flow at high $p_T$~\cite{Andrade:2008xh},
thus improving agreement between hydrodynamics and experimental data. 
Indeed, $v_2$ at high $p_T$ is inferred from azimuthal correlations between 
a high $p_T$ particle and all other particles --- mostly low $p_T$ particles, and 
these azimuthal correlations are reduced due to the decoherence phenomenon. 
Note that the other main explanation for the reduction of $v_2$ at high 
$p_T$, viscosity, typically relies on the assumption of a quadratic momentum dependence of 
the viscous correction to the distribution function at freeze-out $\delta f$, which 
may not be correct~\cite{Dusling:2009df}.

In this paper, we have focused on the transverse momentum dependence 
of the correlations. The rapidity dependence of the correlation is
also worth investigating. In particular, it was recently observed that
azimuthal correlations decrease as a function of the relative 
pseudorapidity~\cite{Pandit:2012rm}, at variance with common lore that
correlations due to flow are essentially independent of rapidity.
While standard models of initial conditions do predict a mild rapidity
dependence of azimuthal correlations~\cite{Bozek:2010vz,Dusling:2009ni}, 
longitudinal fluctuations~\cite{Pang:2012he} could also produce a
decoherence effect similar to the one studied here. 
The detailed structure of two-particle correlations as a function of
both particle momenta thus opens a new window on the study of flow
fluctuations.

\begin{acknowledgments}
We thank Yogiro Hama for useful discussion.  This work is funded by 
FAPESP under projects 09/50180-0 and 09/16860-3, by the FAPESP/CNRS
grant 2011/51854-0, and by CNPq under project  301141/2010-0. ML is
supported by the European Research Council under the 
Advanced Investigator Grant ERC-AD-267258.
\end{acknowledgments}


\begin{thebibliography}{99}

\bibitem{Ackermann:2000tr} 
  K.~H.~Ackermann {\it et al.}  [STAR Collaboration],
  Phys.\ Rev.\ Lett.\  {\bf 86}, 402 (2001)
  [nucl-ex/0009011].
  
\bibitem{Adler:2003kt} 
  S.~S.~Adler {\it et al.}  [PHENIX Collaboration],
  Phys.\ Rev.\ Lett.\  {\bf 91}, 182301 (2003)
  [nucl-ex/0305013].

\bibitem{Back:2004mh} 
  B.~B.~Back {\it et al.}  [PHOBOS Collaboration],
  Phys.\ Rev.\ C {\bf 72}, 051901 (2005)
  [nucl-ex/0407012].
  
\bibitem{Aamodt:2010pa} 
  KAamodt {\it et al.}  [ALICE Collaboration],
  Phys.\ Rev.\ Lett.\  {\bf 105}, 252302 (2010)
  [arXiv:1011.3914 [nucl-ex]].

\bibitem{Ollitrault:1992bk} 
  J.~-Y.~Ollitrault,
  Phys.\ Rev.\ D {\bf 46}, 229 (1992).

\bibitem{Romatschke:2007mq} 
  P.~Romatschke and U.~Romatschke,
  Phys.\ Rev.\ Lett.\  {\bf 99}, 172301 (2007)
  [arXiv:0706.1522 [nucl-th]].

\bibitem{Mishra:2007tw} 
  A.~P.~Mishra, R.~K.~Mohapatra, P.~S.~Saumia and A.~M.~Srivastava,
  Phys.\ Rev.\ C {\bf 77}, 064902 (2008)
  [arXiv:0711.1323 [hep-ph]].

\bibitem{Sorensen:2008dm} 
  P.~Sorensen,
  arXiv:0808.0503 [nucl-ex].
  
\bibitem{Takahashi:2009na} 
  J.~Takahashi, B.~M.~Tavares, W.~L.~Qian, R.~Andrade, F.~Grassi, Y.~Hama, T.~Kodama and N.~Xu,
  Phys.\ Rev.\ Lett.\  {\bf 103}, 242301 (2009)
  [arXiv:0902.4870 [nucl-th]].

\bibitem{Hama:2009vu} 
  Y.~Hama, R.~P.~G.~Andrade, F.~Grassi and W.~-L.~Qian,
  Nonlin.\ Phenom.\ Complex Syst.\  {\bf 12}, 466 (2009)
  [arXiv:0911.0811 [hep-ph]].

\bibitem{Andrade:2010xy} 
  R.~P.~G.~Andrade, F.~Grassi, Y.~Hama and W.~-L.~Qian,
  Phys.\ Lett.\ B {\bf 712}, 226 (2012)
  [arXiv:1008.4612 [nucl-th]].

\bibitem{Sorensen:2010zq} 
  P.~Sorensen,
  J.\ Phys.\ G {\bf 37}, 094011 (2010)
  [arXiv:1002.4878 [nucl-ex]].

\bibitem{Alver:2010gr} 
  B.~Alver and G.~Roland,
  Phys.\ Rev.\ C {\bf 81}, 054905 (2010)
  [Erratum-ibid.\ C {\bf 82}, 039903 (2010)]
  [arXiv:1003.0194 [nucl-th]].

\bibitem{Staig:2010pn} 
  P.~Staig and E.~Shuryak,
  Phys.\ Rev.\ C {\bf 84}, 034908 (2011)
  [arXiv:1008.3139 [nucl-th]];
  Phys.\ Rev.\ C {\bf 84}, 044912 (2011)
  [arXiv:1105.0676 [nucl-th]].

\bibitem{Luzum:2010sp} 
  M.~Luzum,
  Phys.\ Lett.\ B {\bf 696}, 499 (2011)
  [arXiv:1011.5773 [nucl-th]].

\bibitem{Aamodt:2011by} 
  K.~Aamodt {\it et al.}  [ALICE Collaboration],
  Phys.\ Lett.\ B {\bf 708}, 249 (2012)
  [arXiv:1109.2501 [nucl-ex]].

\bibitem{Alver:2010rt} 
  B.~Alver {\it et al.}  [PHOBOS Collaboration],
  Phys.\ Rev.\ C {\bf 81}, 034915 (2010)
  [arXiv:1002.0534 [nucl-ex]].

\bibitem{Chatrchyan:2012wg} 
  S.~Chatrchyan {\it et al.}  [CMS Collaboration],
  Eur.\ Phys.\ J.\ C {\bf 72}, 2012 (2012)
  [arXiv:1201.3158 [nucl-ex]].

\bibitem{ATLAS:2012at} 
  G.~Aad {\it et al.}  [ATLAS Collaboration],
  Phys.\ Rev.\ C {\bf 86}, 014907 (2012)
  [arXiv:1203.3087 [hep-ex]].

\bibitem{Kikola:2011tu} 
  D.~Kikola, L.~Yi, S.~I.~Esumi, F.~Wang and W.~Xie,
  Phys.\ Rev.\ C {\bf 86}, 014901 (2012)
  [arXiv:1110.4809 [nucl-ex]].

\bibitem{Dinh:1999mn} 
  P.~M.~Dinh, N.~Borghini and J.~-Y.~Ollitrault,
  Phys.\ Lett.\ B {\bf 477}, 51 (2000)
  [nucl-th/9912013].

\bibitem{Adare:2011hd} 
  A.~Adare,
  J.\ Phys.\ G {\bf 38}, 124091 (2011)
  [arXiv:1107.0285 [nucl-ex]].

\bibitem{collaboration:2011hfa} 
  J.~Jia,
  J.\ Phys.\ G {\bf 38}, 124012 (2011)
  [arXiv:1107.1468 [nucl-ex]].
  
\bibitem{Miller:2003kd} 
  M.~Miller and R.~Snellings,
  nucl-ex/0312008.

\bibitem{Alver:2006wh} 
  B.~Alver {\it et al.}  [PHOBOS Collaboration],
  Phys.\ Rev.\ Lett.\  {\bf 98}, 242302 (2007)

\bibitem{Luzum:2011mm} 
  M.~Luzum,
  J.\ Phys.\ G {\bf 38}, 124026 (2011)
  [arXiv:1107.0592 [nucl-th]].
  
\bibitem{Ollitrault:2012cm} 
  J.~-Y.~Ollitrault and F.~G.~Gardim,
  arXiv:1210.8345 [nucl-th].

\bibitem{Retinskaya:2012ky} 
  E.~Retinskaya, M.~Luzum and J.~-Y.~Ollitrault,
  Phys.\ Rev.\ Lett.\  {\bf 108}, 252302 (2012)
  [arXiv:1203.0931 [nucl-th]].

\bibitem{Hama:2004rr} 
  Y.~Hama, T.~Kodama and O.~Socolowski, Jr.,
  Braz.\ J.\ Phys.\  {\bf 35}, 24 (2005)
  [hep-ph/0407264].



\bibitem{Drescher:2000ha}
  H.~J.~Drescher, M.~Hladik, S.~Ostapchenko, T.~Pierog and K.~Werner,
  Phys.\ Rept.\  {\bf 350}, 93 (2001)

\bibitem{Gardim:2012yp} 
  F.~G.~Gardim, F.~Grassi, M.~Luzum and J.~-Y.~Ollitrault,
  Phys.\ Rev.\ Lett.\  {\bf 109}, 202302 (2012)
  [arXiv:1203.2882 [nucl-th]].

\bibitem{Gardim:2011qn} 
  F.~G.~Gardim, F.~Grassi, Y.~Hama, M.~Luzum and J.~-Y.~Ollitrault,
  Phys.\ Rev.\ C {\bf 83}, 064901 (2011)
  [arXiv:1103.4605 [nucl-th]].

\bibitem{DerradideSouza:2011rp} 
  R.~D.~de Souza, J.~Takahashi, T.~Kodama and P.~Sorensen,
  Phys.\ Rev.\ C {\bf 85}, 054909 (2012)
  [arXiv:1110.5698 [hep-ph]].

\bibitem{Gardim:2011xv} 
  F.~G.~Gardim, F.~Grassi, M.~Luzum and J.~-Y.~Ollitrault,
  Phys.\ Rev.\ C {\bf 85}, 024908 (2012)
  [arXiv:1111.6538 [nucl-th]].

\bibitem{Floris:2011ru} 
  M.~Floris,
  J.\ Phys.\ G {\bf 38}, 124025 (2011)
  [arXiv:1108.3257 [hep-ex]].

\bibitem{Schenke:2010rr} 
  B.~Schenke, S.~Jeon and C.~Gale,
  Phys.\ Rev.\ Lett.\  {\bf 106}, 042301 (2011)
  [arXiv:1009.3244 [hep-ph]].
%

\bibitem{Niemi:2012aj} 
  H.~Niemi, G.~S.~Denicol, H.~Holopainen and P.~Huovinen,
  arXiv:1212.1008 [nucl-th].

\bibitem{Luzum:2012da} 
  M.~Luzum and J.~-Y.~Ollitrault,
  arXiv:1209.2323 [nucl-ex].

\bibitem{Kapusta:2011gt} 
  J.~I.~Kapusta, B.~Muller and M.~Stephanov,
  Phys.\ Rev.\ C {\bf 85}, 054906 (2012)
  [arXiv:1112.6405 [nucl-th]].

\bibitem{Andrade:2008xh} 
  R.~P.~G.~Andrade, F.~Grassi, Y.~Hama, T.~Kodama and W.~L.~Qian,
  Phys.\ Rev.\ Lett.\  {\bf 101}, 112301 (2008)
  [arXiv:0805.0018 [hep-ph]].

\bibitem{Dusling:2009df} 
  K.~Dusling, G.~D.~Moore and D.~Teaney,
  Phys.\ Rev.\ C {\bf 81}, 034907 (2010)
  [arXiv:0909.0754 [nucl-th]].

\bibitem{Pandit:2012rm} 
  Y.~Pandit [STAR Collaboration],
  arXiv:1209.0244 [nucl-ex].

\bibitem{Bozek:2010vz} 
  P.~Bozek, W.~Broniowski and J.~Moreira,
  Phys.\ Rev.\ C {\bf 83}, 034911 (2011)
  [arXiv:1011.3354 [nucl-th]].

\bibitem{Dusling:2009ni} 
  K.~Dusling, F.~Gelis, T.~Lappi and R.~Venugopalan,
  Nucl.\ Phys.\ A {\bf 836}, 159 (2010)
  [arXiv:0911.2720 [hep-ph]].

\bibitem{Pang:2012he} 
  L.~Pang, Q.~Wang and X.~-N.~Wang,
  Phys.\ Rev.\ C {\bf 86}, 024911 (2012)
  [arXiv:1205.5019 [nucl-th]].

\end{thebibliography}
\end{document}